\documentclass[twocolumn,superscriptaddress]{revtex4-1}

\usepackage{graphicx}
\usepackage{amsmath}
\usepackage{amsfonts}
\usepackage{amssymb}
\usepackage[normalem]{ulem}
\usepackage[svgnames]{xcolor}
\usepackage{bbm}
\usepackage{bm}

\newcommand{\ten}[1]{ \mathfrak{#1} }

\renewcommand{\vec}[1]{ \mathbf{#1} }
\newcommand{\strain}{u}

\begin{document}
\title{Strain-displacement relations and strain engineering in 2d materials}

\author{Daniel Midtvedt}
\affiliation{Max-Planck-Institute for the Physics of Complex Systems, Dresden, Germany}

\author{Caio H. Lewenkopf}
\affiliation{Universidade Federal Fluminense, Niter\'{o}i, Brazil}

\author{Alexander Croy}
\email{croy@pks.mpg.de}
\affiliation{Max-Planck-Institute for the Physics of Complex Systems, Dresden, Germany}

\begin{abstract}
We investigate the electromechanical coupling in 2d materials. For non-Bravais lattices,
we find important corrections to the standard macroscopic strain - microscopic atomic-displacement theory. 
We put forward a general and systematic approach to calculate strain-displacement 
relations for several classes of 2d materials.  
We apply our findings to graphene as a study case, by combining a tight binding and a 
valence force-field model to calculate electronic and mechanical properties of graphene 
nanoribbons under strain. 
The results show good agreement with the predictions of the Dirac equation 
coupled to continuum mechanics. 
For this long wave-limit effective theory, we find that the strain-displacement relations
lead to a renormalization correction to the strain-induced pseudo-magnetic fields.
Implications for nanomechanical properties and electromechanical coupling in 2d 
materials are discussed.
\end{abstract}

\maketitle

\section{Introduction}\label{sec:Intro}
Linear continuum elasticity provides a valuable basis for the investigation
of the mechanical properties of atomic monolayer materials\cite{atis+08,CastroNeto09}.
Elastic theory is also a key element for understanding the relation
between the material deformations and the corresponding modifications 
of its electronic structure\cite{CastroNeto09,Pereira09a,Pereira09b}.
Roughly speaking, strain changes the interatomic distances, thereby 
modulating the overlap of electronic orbitals of neighboring atoms and
modifying the electronic properties of the material. 
To model and engineer this electromechanical coupling it is necessary to 
correctly relate experimentally controllable macroscopic deformations, 
parametrized by the strain tensor $\ten{\strain}$, to microscopic atomic 
displacements. 

We find that for materials with a crystal structure with a basis there is an 
important correction to the standard strain-displacement relations.
We show that by applying strain to a 2d monolayer material, the 
nearest-neighbor vectors connecting the atoms transform as $\vec{r}_{ij}\to 
(\ten{I}_3 + \ten{\strain})\cdot \vec{r}_{ij} + \vec{\Delta}$, where $\ten{I}_n$ is 
the $n\times n$ identity matrix and $\vec{\Delta}$ is a vector that depends on 
the material deformation energy. 

In this Letter we provide a transparent and systematic approach to calculate 
the strain-deformation relations for any kind of 2d material, provided that the 
deformation energy can be parametrized in terms of the bond vectors. 
We show the significance of our findings by contrasting the elastic bulk properties of 
graphene and black phosphorous calculated with and without the proposed 
corrections.

We study the effect of the strain-displacement corrections on 
electronic degrees of freedom in two applications. In the first one 
we incorporate the modified strain-displacement relations in the 
extensively used ${\bf k}\cdot {\bf p}$ theory for graphene, where strain can be 
represented by an effective gauge field\cite{Suzuura2002,Manes2007,Vozmediano10}. 
We show that a correct assessment of $\vec{\Delta}$ renormalizes the strength of 
the gauge vector-potential by a factor $3/5$, which means a factor $(3/5)^2$ for 
scattering rates relevant to transport properties \cite{Couto2014,Burgos2015}. 
A similar correction has previously been identified in the study of the elastic 
properties of carbon nanotubes, where the strain-deformation relations were 
inferred from the analysis of the acoustic phonon modes\cite{woma00,Suzuura2002}. 
In the second application we study the effect of our findings on the gap engineering 
of graphene nanoribbons using strain.

In summary, our study shows that quantitative estimates of the 
electromechanical coupling in 2d materials -- as is often required in strain-engineering 
applications -- requires careful consideration of the lattice deformation on a microscopic level.

\section{Results}
The atomic lattices of 2d materials are characterized by primitive unit cells 
(PUCs)
with a set of basis atoms and two primitive lattice vectors $\vec{a}_1$ and $\vec{a}_2$.
In a homogeneously strained sample it is sufficient to consider the deformation of a 
single unit cell under an applied strain. The deformed primitive lattice vectors read
\begin{equation} 
\label{eq:basis}
    \vec{a}_{i}'=(\ten{I}_3 + \ten{\strain}) \cdot \vec{a}_i,
\end{equation}
where $ \ten{\strain}$ is the strain tensor and $i=1, 2$.
This equation also holds for nonuniform strain, for which 
$\ten{\strain} = \ten{\strain}(\vec{r})$ varies on length scales much longer than the lattice parameter.
In general, the lattice vectors can be expressed by a linear 
combination of the bond vectors $\vec{r}_{jk}$,
\begin{equation}
    \vec{a}_i = \sum_{\vec{r}_{jk}\in \{ \vec{r}_c\}} C_{ijk} \vec{r}_{jk}\;,
\end{equation}
where $\{\vec{r}_c\}$ is the set of bond vectors contained within each unit cell and $C_{ijk}$ 
is a tensor encoding the lattice connectivity. We address the case where the strain field 
deforms the lattice, but preserves its connectivity. 
We show that for materials with a non-Bravais lattice structure, the modifications
of the bond vectors due to strain do not follow Eq.~(\ref{eq:basis}). Let us start from the 
most general relation
\begin{equation}
    \vec{r}_{ij}'=(\ten{I}_3 + \ten{\strain})\cdot \vec{r}_{ij} + \vec{\Delta}_{ij}\;,
\end{equation}
where the vectors $\vec{\Delta}_{ij}$ contain the differences between the strain
displacement relations for $\vec{a}_i$ and $\vec{r}_{ij}$. 
The number of independent vectors $\vec{\Delta}_{ij}$ is determined by the number of basis 
atoms and $\sum_{\vec{r}_{jk}\in \{ \vec{r}\}_c} C_{ijk} \vec{\Delta}_{jk}=0$. 
For a Bravais lattice $\vec{\Delta}_{ij}=0$. 
Each additional atom introduces three degrees of freedom, that can be expressed by an
in-plane vector $\vec{\Delta}_\parallel$ and an out-of-plane component $\Delta_\perp\, \hat{z}$. 
The number of independent displacement vectors obtained in this way can, in general, be 
reduced by enforcing the lattice symmetry and connectivity. 
To be specific, we consider the most important 2d materials\cite{miau+14} under current 
investigation, divided in three groups, shown in Fig.\ \ref{fig:Lat}. 

The first class contains graphene and graphene-like materials, such as boron nitride, 
as well as non-planar materials, such as silicene and germanene. 
The PUC consists of two basis atoms and two lattice vectors $\vec{a}_1=
a_0 \left(3/2,\sqrt{3}/2,0\right)$ and $\vec{a}_2=a_0 \left(3/2,-\sqrt{3}/2,0\right)$,
see Fig.\ \ref{fig:Lat}(a).
The three nearest neighbor vectors are given by $\vec{r}_{1}= \left(-a_0,0,0\right) - 
h\,\hat{z}$, $\vec{r}_{2}=\vec{a}_1 + \vec{r}_1$ and $\vec{r}_{3}=\vec{a}_2 + \vec{r}_1$, where 
$h$ is the PUC corrugation height. 
The lattice vectors are written in terms of the bond vectors as 
$\vec{a}_1 = \vec{r}_{2}-\vec{r}_1$ and $\vec{a}_2 =\vec{r}_{3}-\vec{r}_1$. 
The bond vectors transform as $\vec{r}_{i}'=\vec{r}_{i} + \ten{\strain}\cdot \vec{r}_{i} + \vec{\Delta}_{i}$, 
with the constraints $0=\vec{\Delta}_2-\vec{\Delta}_1$ and $0=\vec{\Delta}_3-\vec{\Delta}_1$. 
Hence, $\vec{\Delta}_i\equiv \vec{\Delta}$. The latter is conveniently expressed by in-plane 
deformations $\vec{\Delta}_{\parallel}$ and the corrugation change $\Delta_\perp \hat{z}$. 
Graphene is a special case for which $\Delta_\perp\equiv 0$. 

The second class contains the transition metal dichalcogenides (TMDCs). Here the PUC contains 
three atoms, one transition metal (${\rm Mo}$ or ${\rm W}$) and two chalcogens (${\rm S}$ or 
${\rm Se}$), see Fig.~\ref{fig:Lat}(b). The transition metal has six nearest neighbors. The bond 
vectors are labeled $\vec{r}_{1,u/d}$, $\vec{r}_{2,u/d}$ and $\vec{r}_{3,u/d}$ where the subscript 
$u$ or $d$ indicates whether the bond connects to the upper ($u$) or lower ($d$) layer of chalcogen 
atoms. The lattice vectors are given by $\vec{a}_1 = \vec{r}_{2,u/d}-\vec{r}_{1,u/d}$ and 
$\vec{a}_2 = \vec{r}_{3,u/d}-\vec{r}_{1,u/d1}$. The bond vectors transform as 
$\vec{r}_{i,u/d}'=\vec{r}_{i,u/d} + \ten{\strain}\cdot \vec{r}_{i} + \vec{\Delta}_{i,\parallel} \pm 
\Delta_{i,\perp}/2\hat{z}$. The vectors $\vec{\Delta}_{i,\parallel}$ now satisfy 
$0=\vec{\Delta}_{2,\parallel}-\vec{\Delta}_{1,\parallel}$ and 
$0=\vec{\Delta}_{3,\parallel}-\vec{\Delta}_{1,\parallel}$. 
Consequently $\vec{\Delta}_{i,\parallel}\equiv \vec{\Delta}_{\parallel}$. 
By inversion symmetry, vectors connecting to the upper layer and the lower layer transform in 
the same way, but with opposite signs in the out-of-plane direction. Hence, the strain-displacement 
response of TMDCs is also characterized by a single in-plane vector $\vec{\Delta}_{\parallel}$ 
and the change in inter-chalcogen distance $\Delta_\perp$. 

The third class consists of a single material, phosphorene, an atomically thin puckered 
material derived from layered black phosphorus. Its PUC contains four atoms and six 
unique bond vectors, see Fig.~\ref{fig:Lat}(c). Four bond vectors are situated in the puckers 
and are denoted by $\vec{r}_{i,u/d}$ with $i=1,\;2$ and subscripts $u/d$. The two remaining 
vectors connect the upper and lower puckers, and are labeled $\vec{r}_{cc}$ and $\vec{r}_{cc'}$. 
We find that $\vec{a}_1 = -\vec{r}_{cc} + \vec{r}_{1,u} + \vec{r}_{cc'} + \vec{r}_{2,d}$, 
and $\vec{a}_2 = -\vec{r}_{2,u} + \vec{r}_{1,u}$. 
Alternatively, we also write $\vec{a}_1 = -\vec{r}_{cc} + \vec{r}_{2,u} + \vec{r}_{cc}' + \vec{r}_{1,d}$ 
and $\vec{a}_2 = -\vec{r}_{2,d} + \vec{r}_{1,d}$. Hence, we obtain the constraints $0 = -\vec{\Delta}_{cc} + \vec{\Delta}_{1,u} + \vec{\Delta}_{cc'} + \vec{\Delta}_{2,d}$, $0 = -\vec{r}_{2,u} + \vec{r}_{1,u}$. 
This leads to four equations for the vectors $\vec{\Delta}_i$. Using inversion symmetry 
and the same arguments as for the TMDCs, the strain displacement relation in 
phosphorene is also characterized by a single in-plane vector $\vec{\Delta}_{\parallel}$ 
and a change in vertical inter-pucker distance $\Delta_\perp$. 

We conclude that upon application of strain to all these 2d materials, due to lattice connectivity and symmetry constrains,
the strain-displacement correction is simply
\begin{equation}\label{eq:transf}
    \vec{\Delta}_{ij} = \vec{\Delta}_\parallel + \Delta_\perp \hat{z},
\end{equation}
independent of $i$ and $j$.
Both $\vec{\Delta}_\parallel$ and $\Delta_\perp$ depend on the applied strain and on 
the interatomic interactions. Their calculation requires a microscopic model to 
account for the deformation energy. For uniform strain, one can determine the 
strain-displacement correction using the geometric considerations of the
previous paragraph and, for instance, first principles calculations. For non-uniform strain 
and/or finite lattices, this procedure becomes computationally prohibitive, and one has 
to resort to semi-empirical force-field models.

\begin{figure*}[ht!]
  \centering
  \includegraphics[width=0.9\textwidth]
             {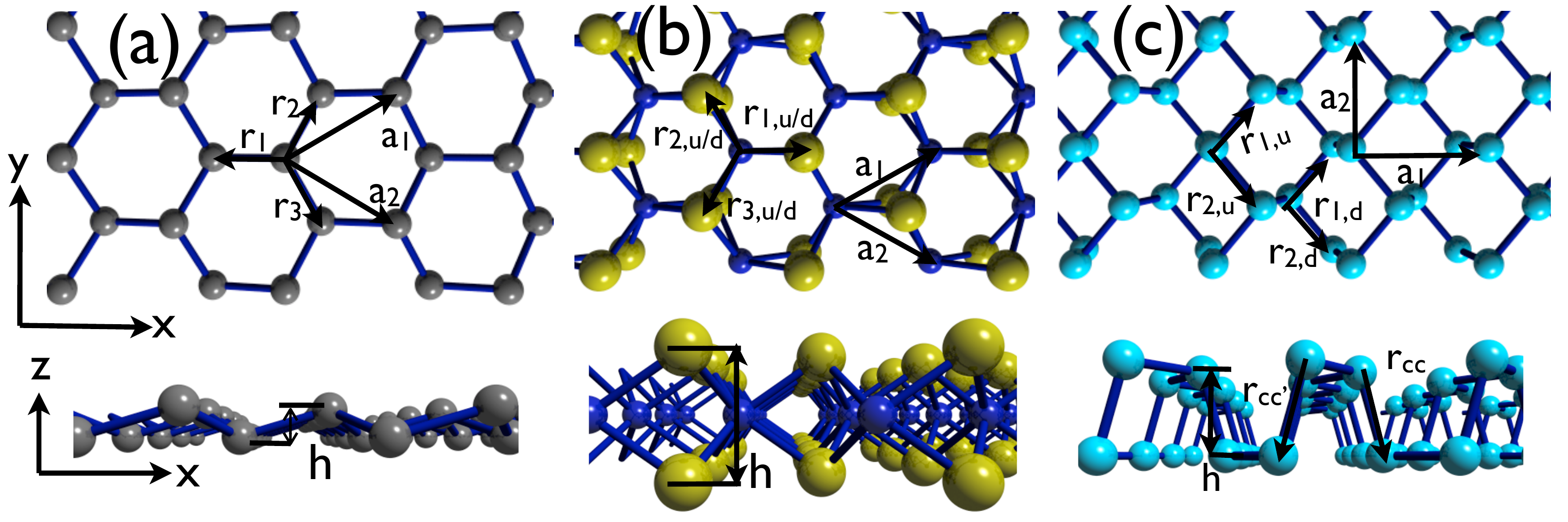}           
             \caption{Lattice structures of the three classes of 2d materials considered. a) The unit cell of graphene-like materials are defined by two lattice vectors $\vec{a}_1$ and $\vec{a}_2$ and three nearest neighbor vectors $\vec{r}_1$, $\vec{r}_2$ and $\vec{r}_3$. The material can in general be corrugated, signified by $h$ in the lower figure. Graphene is contained as the case $h=0$. b) The unit cell of TMDCs consist of two lattice vectors and three basis atoms, which define six nearest neighbor vectors. Due to the layered structure of the material, it has an effective thickness $h$. c) The phosphorene lattice is characterized by two lattice vectors and four basis atoms, defining ten nearest neighbor vectors. Its puckered structure gives rise to the effective thickness $h$.}
\label{fig:Lat}
\end{figure*}

In the following, we put forward a systematic and transparent approach to calculate
$\vec{\Delta}_\parallel$ and $\Delta_\perp$ for any 2d material whose deformation 
energy is described by a force-field model parameterized by the bond vectors.
Valence force models (VFMs) are a convenient choice since they offer good accuracy 
at low computational costs. There are various VFMs for the different 2d materials 
\cite{kaka+86,wasm+75,pete09,jipa+13,ji15}. 
Let us focus on graphene-like materials, which can be accurately 
addressed using the VFM introduced by Perebeinos and Tersoff\cite{pete09} to 
describe the interactions between $sp^2$-bonded carbon atoms. 
The deformation energy is given by\cite{pete09}
\begin{align}
\label{eq:PeTe}
	E_{\rm def}={}& \frac{\beta_r}{ a_0^2} \sum_{i,j\in i} \left(\delta r_{ij}\right)^2 + \beta_c\sum_{i,j<k\in i} (\delta c_{i,jk})^2 \nonumber \\
	    &+\frac{\beta_{r2}}{ a_0^{2}} \sum_{i,j<k\in i} \left(\delta r_{ij}\right) \left(\delta r_{ik}\right) \nonumber\\
&+ \frac{\beta_{rc}}{ a_0}\sum_{i,j\neq k<l\in i} \left(\delta r_{ij}\right)\left(\delta c_{i,kl}\right) \nonumber\\
&+ \frac{\beta_v}{a_0} \sum_{i,j<k<l\in i} \left(\frac{3 \vec{r}_{ij}'\cdot\vec{r}_{ik}'\times\vec{r}_{il}'}{r_{ij}r_{ik}+r_{ik}r_{il}+r_{il}r_{ij}}\right)^2\;,
\end{align}
where $r_{ij}=|\vec{r}_{ij}'|$ is the bond length, $\delta r_{ij}=r_{ij}'-a_0$ is the 
change in bond length and $\delta c_{i,jk}$ is defined as $\delta c_{i,jk}=
\cos \theta_{i,jk}'-\cos \theta_{i,jk}$. Here $\theta_{i,jk}$ is the angle between 
atoms $i$, $j$ and $k$ with atom $i$ as apex in equilibrium, while $\theta_{i,jk}'$ 
is the angle in the deformed lattice. 
The summations in Eq.\ \eqref{eq:PeTe} follow the convention: 
(a) $j\in i$ indicates that the index $j$ runs over the three neighbors of atom $i$; 
(b) for $j<k\in i$ both $j$ and $k$ are neighbors of the atom $i$, and are ordered 
to avoid double counting, leaving three possible terms; (c) $j\neq k<l\in i$ leaves three terms for each $i$. 
The first and second terms in Eq.\ \eqref{eq:PeTe} give the energy cost of stretching 
and bending bonds as in a Keating model\cite{ke66}. The third and the fourth terms 
couple stretching of different bonds and couple stretching and bending, respectively. 
The last term is related to out-of-plane displacements.
In Ref.\ \onlinecite{pete09}, there is an additional term which penalizes misalignments 
of neighboring $\pi$-orbitals that we neglect here.

For a given strain, we minimize $E_{\rm def}$ with respect to $\vec{\Delta}_\parallel$ and 
$\Delta_\perp \hat{z}$ to obtain
\begin{subequations}\label{eq:DELTA}
\begin{align}
    \vec{\Delta}_\parallel \equiv{}& -\frac{a_0 \kappa(h)}{2} \left(
    \begin{matrix}
        \strain_{yy}-\strain_{xx}\\
        2\strain_{xy} \\
        0
    \end{matrix}
    \right)\;, \\
    \Delta_\perp \equiv{}& \pm {a_0 \kappa_{\perp}(h)} (\strain_{xx} + \strain_{yy}) \,,
\end{align}
\end{subequations}
where $\kappa(h)$ and $\kappa_{\perp}(h)$ are functions of the unit cell corrugation 
height $h$ and characterize the changes in the PUC internal structure, in addition to 
the trivial shift of atomic positions when the PUC is strained.
To lowest order in $h/a_0$,
\begin{subequations}\label{eq:kappa}
    \begin{align}
        \kappa(h) \approx{}& -\frac{9\beta_c-4\beta_r+2\beta_{r2}}{9\beta_c + 4\beta_r-2(\beta_{r2}+3\beta_{rc})}
        \\
    \kappa_{\perp}(h) \approx{}& - \frac{h}{a_0} \frac{2\beta_r + 2\beta_{r2}+3\beta_{rc}}{9\beta_v}
   \;.
\end{align}
\end{subequations}

Upon elimination of $\vec{\Delta}_\parallel$ and $\Delta_\perp$, 
the microscopic deformation energy per unit area, $\mathcal{E}=E_{\rm def}/{\cal A}$, 
becomes the standard expression for the continuum elastic energy-density of an 
isotropic membrane\cite{lali86},
\begin{align} 
    \mathcal{E}={}&\frac{1}{2}\lambda (\strain_{xx}+\strain_{yy})^2 + \mu (\strain_{xx}^2+\strain_{yy}^2+2\strain_{xy}^2)\;\nonumber\\
    ={}&\frac{Y_{2d}}{2} (\strain_{xx}^2+\strain_{yy}^2 +2\nu \strain_{xx}\strain_{yy} + (1-\nu) \strain_{xy}^2),\label{eq:cm_energy}
\end{align}
reconciling our results with previous works on the long-wavelength elastic behavior 
of graphene\cite{atis+08,lewe+08,zhak+11}. Here, $\lambda$ and $\mu$ are the 
Lam\'{e} parameters of the material, and $Y_{2d}$ and $\nu$ are the $2d$ Young 
modulus and Poisson ratio, respectively. They are related via
\begin{equation}\label{eq:Young_Poisson}
	Y_{{\rm 2d}} = \lambda + 2\mu\;,\quad\text{and}\quad
	\nu = \frac{\lambda}{\lambda + 2\mu}\;.
\end{equation}
We find that the Lam\'{e} parameters, given in terms of the microscopic parameters 
of Eq.~\eqref{eq:PeTe}, are
\begin{subequations}\label{eq:Lame}
\begin{align}
    \lambda=& \frac{1}{\sqrt{3} a_0^2}\left[\frac{ 8 \beta_r^2- 18 \beta_c \beta_r+4\beta_r\left(\beta_{r2}-3\beta_{rc}\right)}{9\beta_{c} + 4\beta_r - 2\left(\beta_{r2}+3\beta_{rc}\right)}\right.\nonumber\\
    &\left.+\frac{9\beta_{rc}^2-36\beta_c \beta_{r2} - 4\beta_{r2}^2  -12\beta_{rc}\beta_{r2} }{9\beta_{c} + 4\beta_r - 2\left(\beta_{r2}+3\beta_{rc}\right)} \right] \;,  \\
    \mu=&\frac{3 \sqrt{3}}{a_0^2}\left[\frac{4\beta_c \beta_r - 2\beta_c \beta_{r2} - \beta_{rc}^2}{9\beta_{c} + 4\beta_r - 2\left(\beta_{r2}+3\beta_{rc}\right)}\right]\;.
\end{align}
\end{subequations}
For graphene, the parameter values reported in Ref.\ \onlinecite{pete09} together with 
Eqs.\ \eqref{eq:kappa} lead to $\kappa=0.39$ and $\kappa_{\perp}=0$.
Moreover, Eqs.\ \eqref{eq:Lame} give $\lambda=4.4$ eV/{\AA}$^2$ and $\mu=8.8$\, eV/{\AA}$^2$, which implies $Y_{{\rm 2d}}=352\,{\rm N/m}$ and $\nu=0.2$.

Neglecting the bond-bond and bond-angle correlations in Eq.\ \eqref{eq:PeTe} 
(setting $\beta_{r2}=\beta_{rc}=0$), we find that $\kappa(h)$ depends only on 
the Poisson ratio, $\kappa(h)\approx 2\nu/(1+\nu)$. Consequently, $\kappa(0)$ 
vanishes for materials with a negligible Poisson ratio (where $\mu\gg \lambda$), 
and is limited from above by $\kappa(0)<2/3$ for isotropic materials (for which \cite{lali86}
$\nu<1/2$). In this simplified VFM, $\kappa(0)$ agrees with the findings 
of Ref.~\onlinecite{Suzuura2002}, and may serve as a rough estimate, since it can be 
readily obtained form the Poisson ratio which is a macroscopically observable quantity. 
For $\nu=0.2$ this approximation gives $\kappa\approx 1/3$, 
which agrees reasonably well with the value obtained from the full set of parameters ($\kappa=0.39$). 
Similarly, we find $\kappa_\perp(h) \approx - 2 (h/a_0) (\beta_r/(9\beta_v))$.

The approximation $\kappa\approx 2\nu/(1+\nu)$ is expected to hold also for TMDCs, due to the hexagonal structure of the lattice when projected onto the monolayer plane. For example, using $\nu_{\rm MoS_2}=0.27$ \cite{fe76}, we obtain $\kappa_{\rm MoS_2}\approx 0.43$, which is slightly larger than the estimate for graphene. 
For phosphorene, the strain-displacement relation no longer obeys the simple form of 
Eq.\ \eqref{eq:DELTA},  
since $\vec{\Delta}$ shows a directionality reflecting the material anisotropy. 
Nonetheless, we can still use our theory to estimate the elastic properties of phosphorene (see Supplementary Material). 
Using a VFM developed for layered black phosphorus \cite{kaka+82}, we obtain sound velocities
$v_{xx}=3508$ m/s, $v_{yy}=8147$ m/s, and $v_{xy}=3707$ m/s in good agreement with 
Ref.~\onlinecite{kaka+82} (without correction $v_{xx}$ and $v_{xy}$ are overestimated by a factor 2). The Young 
moduli are $Y_x = 17$ N/m and $Y_y =94$ N/m (without corrections $Y_x = 58$ N/m and 
$Y_y =95$ N/m). These values are in good agreement with DFT calculations \cite{wepe14,qiko+14,elkh+15}, where
$Y_x = 24 \ldots 29$ N/m and $Y_y=88 \ldots 102$ N/m. 
Hence, a correct treatment of the strain-displacement relations is necessary to account for the predicted anisotropy of phosphorene\cite{jipa14}.

\begin{figure}[t]
  \centering
  \includegraphics[width=0.48\textwidth]
             {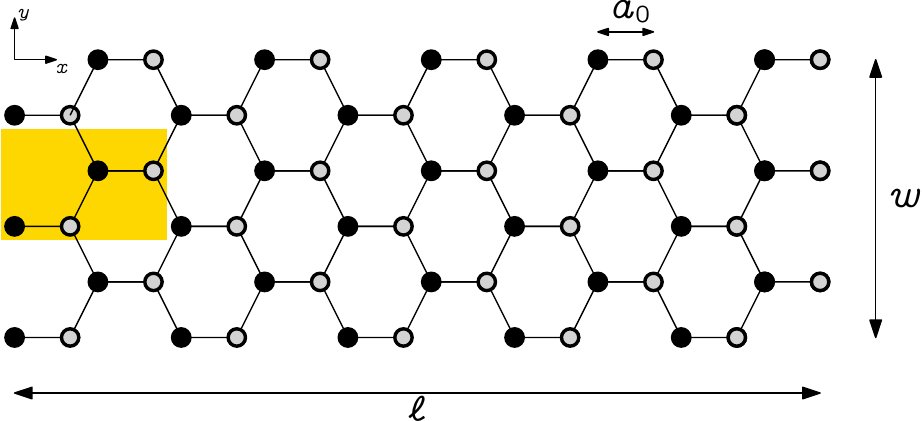}           
\caption{Structure of an armchair nanoribbon. The highlighted area 
represents the system unit cell. The ribbon 
consists of $N_1 \times N_2$ cells, where $N_1$ counts the cells in 
$x$-direction and $N_2$ in $y$-direction. In the lateral direction 
periodic boundary conditions are imposed at the edges.}
\label{fig:ribbon}
\end{figure}
%
\begin{figure}[htb]
  \centering
  \includegraphics[height=0.39\textwidth]
             {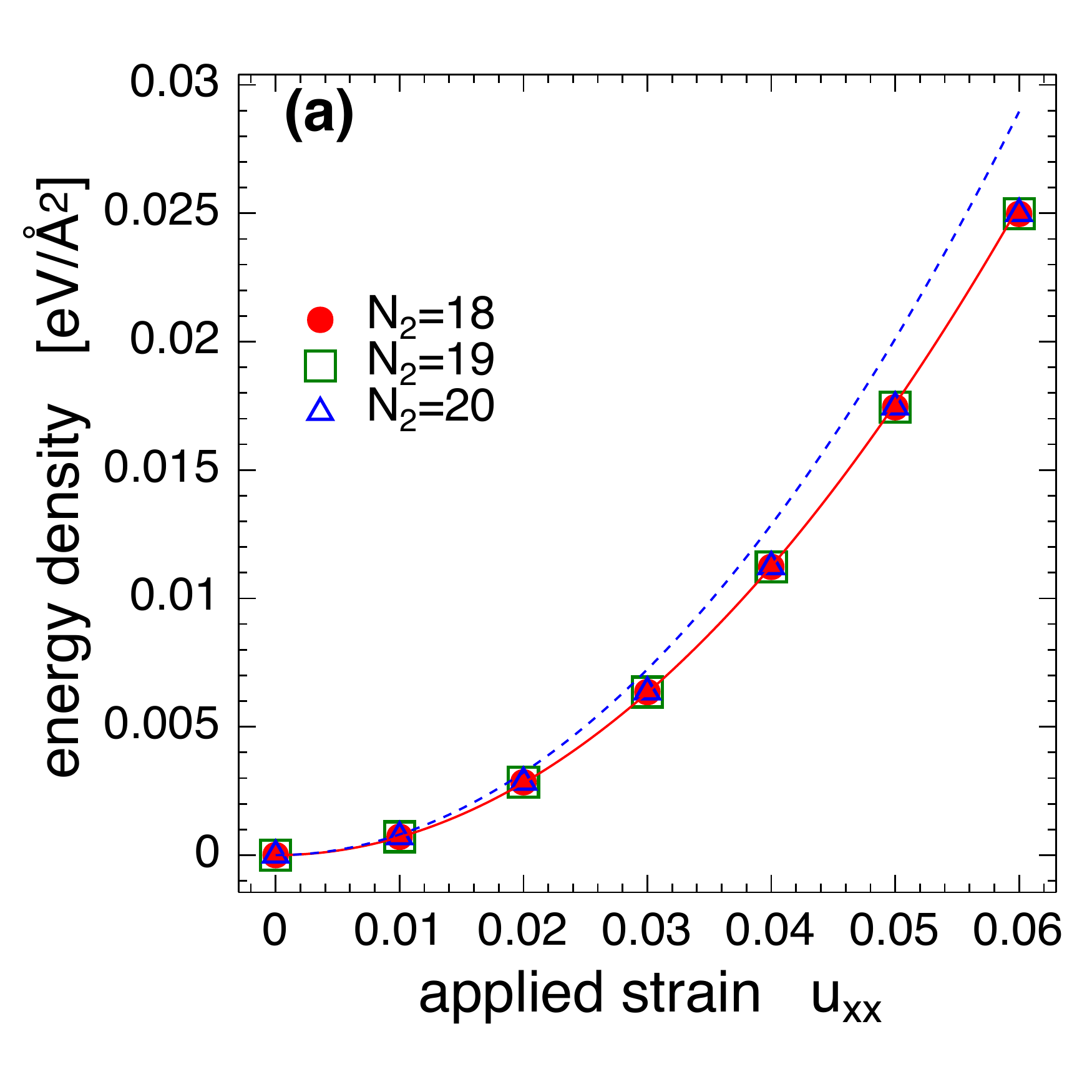}
  \includegraphics[height=0.40\textwidth]
             {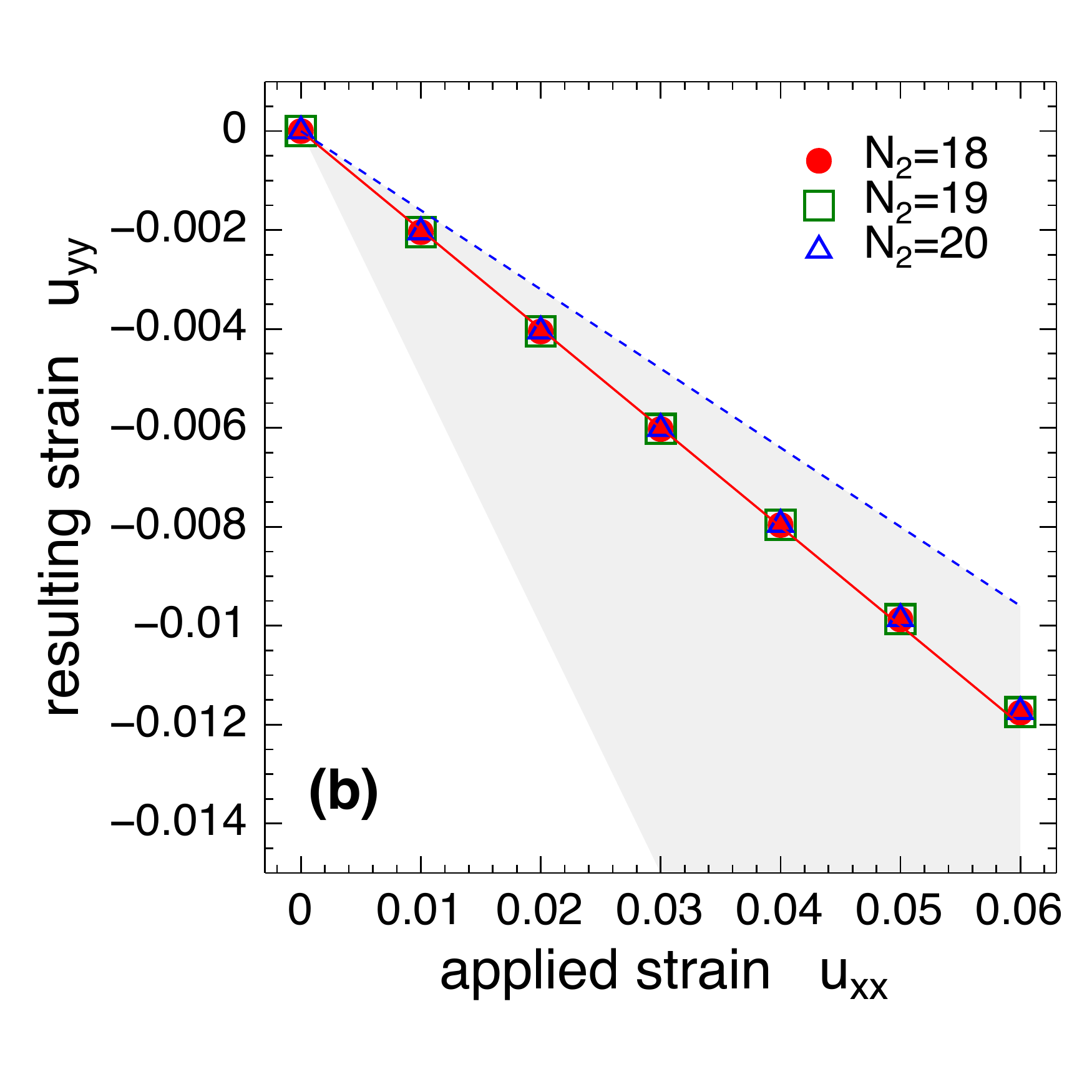}
    \caption{Strain $u_{xx}$ dependence of (a) the elastic energy and (b) the resulting 
                  strain $u_{yy}$ 
                   for AGNRs of different widths ($N_1=300$ in all cases). 
	          Symbols denote numerical  results. The full (dashed) line stands for 
	          the continuum model results with $\kappa\approx0.39$ ($\kappa=0$). The shaded region
          indicates the range of Poisson ratios from $\nu(\kappa=0)=0.16$ to $\nu=1/2$.}
  \label{fig:CM}
\end{figure}
We illustrate our results by studying armchair graphene nanoribbons (AGNRs) 
stretched along the longitudinal direction. See Fig.\ \ref{fig:ribbon} for a sketch of the 
setup. The equilibrium configuration is obtained by minimizing $E_{\rm def}$, 
Eq.\ \eqref{eq:PeTe}. 
The mechanical energy density as a function of the applied strain $u_{xx}$ is 
shown in Fig.\ \ref{fig:CM}(a). 
Numerical results are denoted by symbols, whereas full and dashed lines correspond 
to analytical results obtained from Eq.\ \eqref{eq:cm_energy} for $\kappa=0.39$ and $\kappa=0$.
By setting $\kappa=0$, one obtains a slightly larger value of $Y_{2d}$ such that 
the stretching energy is overestimated. 
We also calculate the resulting strain $u_{yy}$ in the transversal direction. As the 
ribbon contracts upon stretching, $u_{yy}$ is negative for $u_{xx}>0$ and decreases 
with increasing strain. This is shown in Fig.\ \ref{fig:CM}(b). The slope of the curve 
at $u_{xx}=0$ gives the Poisson ratio, which agrees very well with the estimate obtained 
from Eq.\ \eqref{eq:Young_Poisson}. The standard approach, for which $\kappa=0$,
gives $\nu(\kappa=0)\approx0.16$, thus underestimating the Poisson ratio.

The strain-displacement relations presented above significantly modify the electronic 
properties of deformed 2d materials, which are crucial for strain-engineering\cite{Pereira09a,Pereira09b,lugu10}.
Essentially, mechanical deformations have two effects on the electrons. 
Firstly, the change in electron-ion potential in the neighborhood of an atom generates 
an on-site potential often referred to as the deformation potential. Secondly, changes 
in the distances between neighboring atoms modify the overlaps of the corresponding 
orbital wave-functions and thus the electronic structure. 

These effects are accounted for in the nearest-neighbor hopping Hamiltonian which 
provides an accurate description of the low-energy $\pi$-bands in 
graphene\cite{Suzuura2002,CastroNeto09,Vozmediano10}, namely,
\begin{multline}
	H = \sum_{i=1}^{N_{\rm at}} \left[ \left(3 v_0(a_0) + \frac{g}{a_0} \sum_{<k,i>}\frac{\vec{r}_{ik} \cdot \delta\vec{ r}_{ik}}{a_0} \right) c^\dagger_i c_i^{} \right.\\\left.
	- t_{0}(a_0) \sum_{<j,i>} \left(1 - \frac{\beta}{a_0}\frac{\vec{r}_{ij} \cdot \delta\vec{ r}_{ij}}{a_0} \right)c^\dagger_i c_j^{} \right] \;. \label{eq:TBHamiltonian}
\end{multline}
Here $\beta/a_0 = -{t_{0}'(a_0)}/{t_{0}(a_0)}$, $g/a_0 = v_{0}'(a_0)$, where $t_0(r)$ is the 
(distance dependent) hopping amplitude and $v_0(r)$ is the electron-ion potential. In the following
we omit the overall on-site energy $3v_0(a_0)$. 

To obtain analytical insight, it is customary \cite{CastroNeto09,Vozmediano10,Couto2014,Burgos2015} 
to consider 
the low-energy limit of the tight-binding model above. By expressing Eq.\ \eqref{eq:TBHamiltonian} 
in reciprocal space and taking its long wavelength limit, that is, by expanding $H$ to linear 
order in momentum around the $\vec{K}$ and $\vec{K}'$-points \cite{CastroNeto09,Vozmediano10},
one obtains an effective Dirac Hamiltonian. 
For $\vec{K}$ we write
\begin{equation}
    H_{\vec{K}} = v_F \bm{\sigma} \cdot ({\bf p} - {\bf A}) + v_{\rm D} \ten{I}_2\;,
\end{equation}
where $v_F = 3a_0t_0/2\approx 10^5$ m/s, $\bm{\sigma}$ are Pauli matrices, 
${\bf p}$ is the momentum, $\vec{A}$ is the vector potential, and $v_{{\rm D}}$ is 
the scalar deformation potential. The latter are given by
\begin{subequations}
\begin{align}
    A_x ={}& \frac{\beta (1-\kappa)}{a_0} u_{xy}\;\\
    A_y ={}& \frac{\beta (1-\kappa)}{2 a_0} \left( u_{yy} - u_{xx} \right)\;,\\
    v_{\rm D} ={}& \frac{3 g}{2} \left[1 + 2\kappa_{\perp}(h) h\right] \left( u_{xx} + u_{yy} \right)\;.
\end{align}
\end{subequations}

Hence, we find that the strain-displacement relations 
renormalize $\vec{A}$ by a factor $1-\kappa$ with respect to the standard elasticity-induced gauge 
theory \cite{Vozmediano10,Couto2014,Guinea2010,Levy2010,Low2011,deJuan12}.
For graphene, the material parameters\cite{pete09} give $1-\kappa \approx 3/5$. 
(Using the approximation $\kappa\approx{2\nu}/(1+\nu)$ we obtain
$1-\kappa\approx (1-\nu)/(1+\nu)=2/3$.)
Based on a VFM that drastically overestimates the Poisson ratio, previous works\cite{woma00,Suzuura2002} 
report a reduction factor of about $1/3$. Note
that in the derivation above, we ignore the strain-dependent renormalization of the Fermi velocity reported 
previously\cite{deJuan12}, since it has a negligible effect on our results.

The reduced vector potential has an important effect on strain engineering 
applications\cite{Guinea2010,Low2011}. Let us analyze, for instance,  the 
first (sub)band-gap 
of AGNRs\cite{lugu10}. In the long wavelength limit, the 
magnitude of this band gap is given by
\begin{multline}\label{eq:gap}
    E_g\approx \sqrt{3} |t_0| \min_{n=0,1,\ldots}\left| \frac{\pi}{N_2+1} (n - \phi) \right.\\\left. 
           - \frac{\sqrt{3}\beta_D}{2}(1-\kappa)(1+\nu)u_{xx}\right|
\end{multline}
with $\phi=0$ for metallic and $\phi=1/3$ for semiconducting AGNRs and where 
$N_2$ is the number of unit cells in the transversal direction and $n$ is the band 
index. The magnitude of the band-gap as a function of strain is piecewise linear 
and shows a characteristic zig-zag shape. We predict that the slope of 
the linear regions is decreased by a factor $(1-\kappa)$($\approx 3/5$ for graphene) due 
to the strain-displacement relations, while the positions of the maxima and minima of 
the band gap are shifted by a factor $1/(1-\kappa)$($\approx 5/3$ for graphene). 

\begin{figure*}[t]
  \centering
  \includegraphics[height=0.33\textwidth]
  {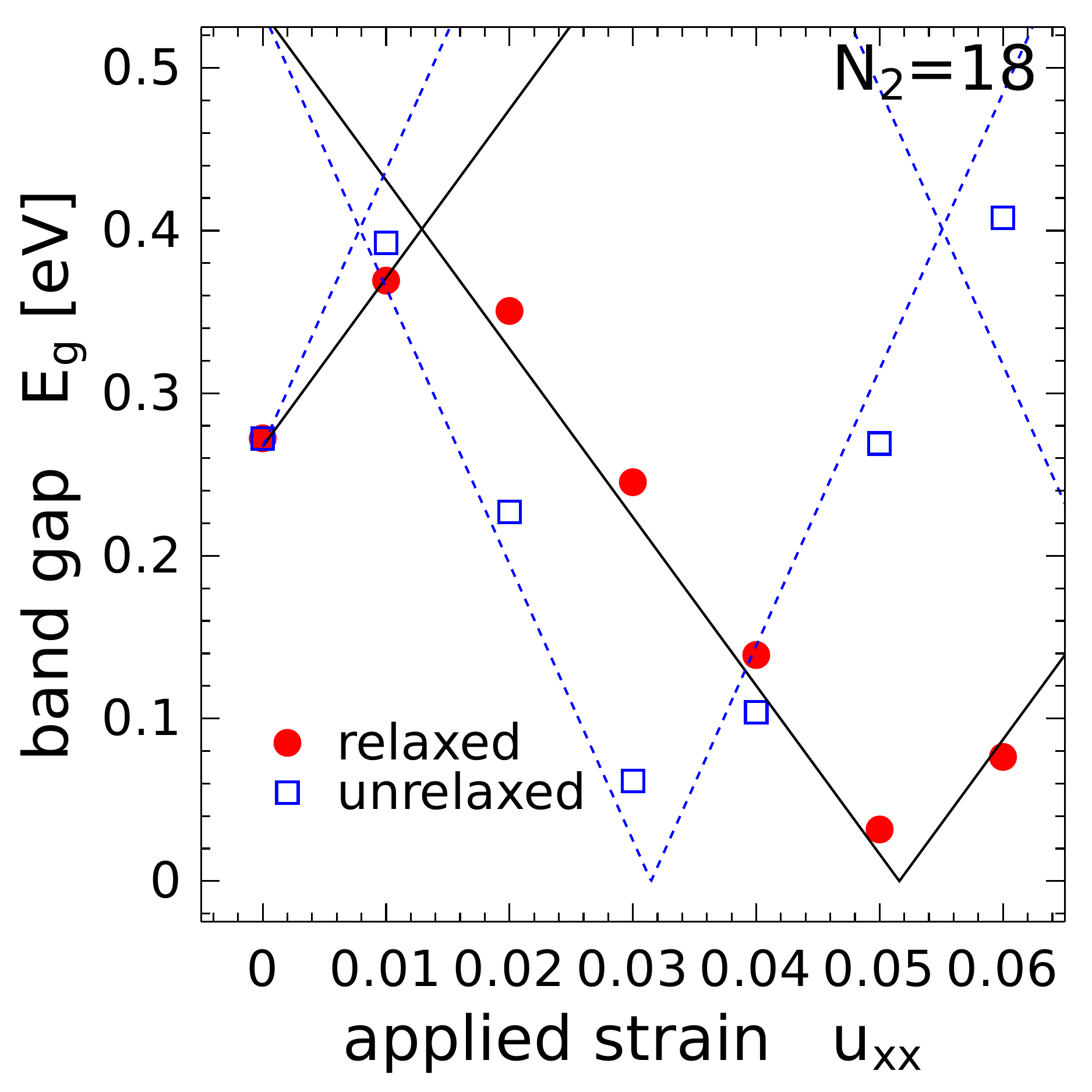}\hspace{1cm}
  \includegraphics[height=0.33\textwidth]
             {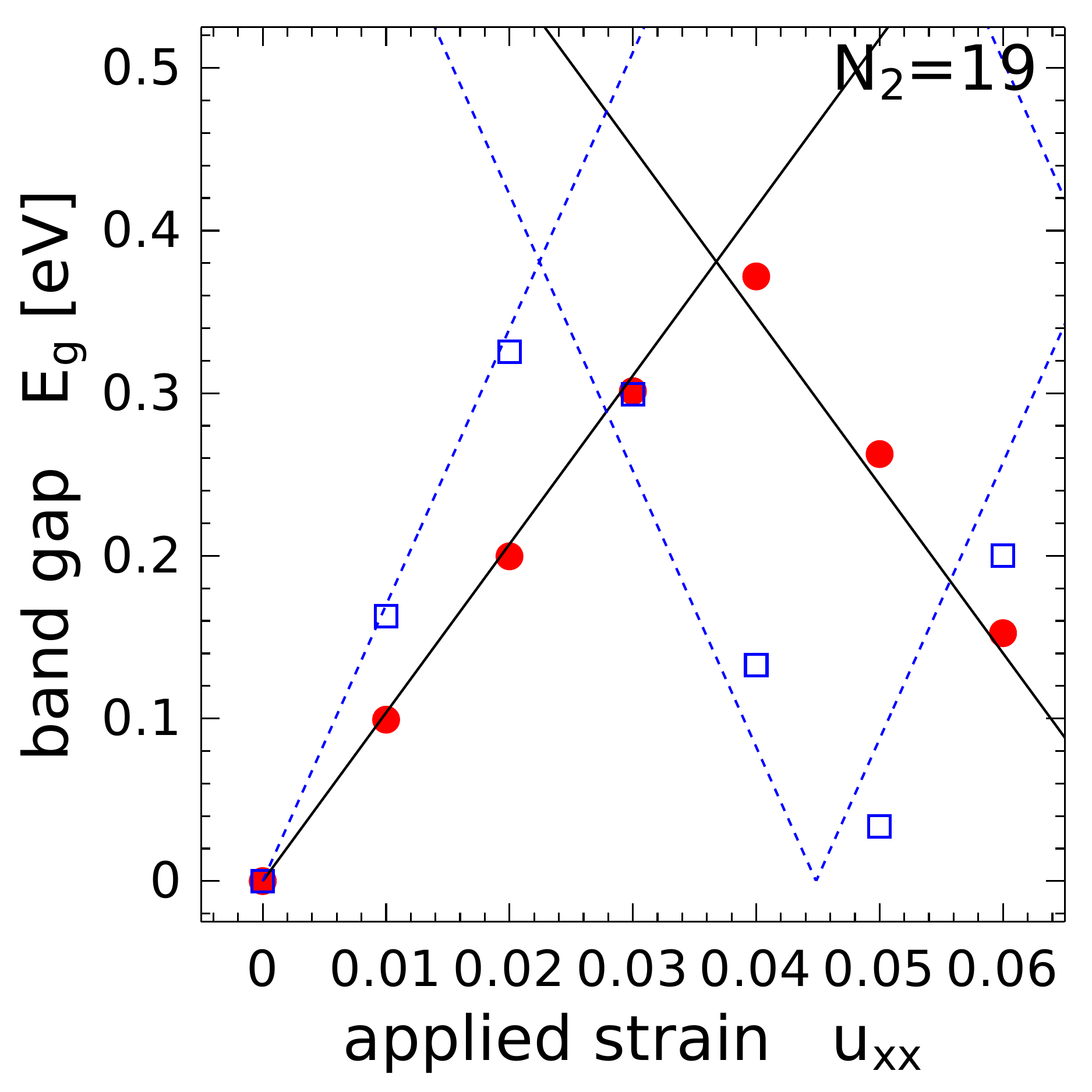}
\caption{Strain dependence of the (first) gap size for AGNRs of different widths ($N_1=300$
                 in all cases). Symbols represent tight-binding calculations without (blue) and with 
                 relaxation (red). The dashed (solid) lines show the effective Dirac equation, 
                 Eq.\ \eqref{eq:gap}, results for $\kappa=0$ ($\kappa\approx0.39$).}
  \label{fig:Gap}
\end{figure*}

We verify this prediction numerically. First, we obtain the relaxed lattice structure 
of a stretched AGNR as described previously. Subsequently, we diagonalize the 
Hamiltonian \eqref{eq:TBHamiltonian} using $t_0=2.8\;\text{eV}$, $g=4\;\text{eV}$, 
$\beta=3.37$\cite{CastroNeto09,Pereira09a}, and the relaxed lattice structure as an 
input. From the eigenvalues we 
directly calculate the band-gap $E_{\rm g}$ shown in Fig.\ \ref{fig:Gap}. 
For comparison we also show $E_{\rm g}$ for $\kappa=0$. 
For the considered range of applied strain we find a good agreement between the 
numerical results and Eq.\ \eqref{eq:gap}. 
We observed that the strain required to achieve a certain gap size increases for $\kappa>0$. 

A similar renormalization is also found for phosphorene.
Based on the two orbital tight-binding model put forward in Ref.~\onlinecite{jipa15}
and the strain-displacement relations derived in the supplementary material we find
an \emph{anisotropic} renormalization of the strain-induced band-gap. It is reduced
by factors $2.5$ and $1.35$ when straining in $x$ and $y$ direction, respectively.

\section{Conclusions}
Two-dimensional materials have gained considerable interest, in particular due to 
the enticing prospect of engineering their electronic properties using strain. Our results 
show that the relationship between 
strain and electronic properties is less trivial than often assumed. 
For Bravais lattices, the relation between strain and displacements is simple, namely the 
bond vectors transform as $\vec{r}_{ij}\to (\ten{I}_3+\ten{\strain})\cdot\vec{r}_{ij}$. For 
non-Bravais lattices, the basis atoms introduce additional degrees of freedom, significantly 
modifying the relation between strain and displacement. 
With few exceptions\cite{woma00,Suzuura2002}, this fact is typically neglected in the literature 
on electronic properties of deformed 2d materials
\cite{Manes2007,Pereira09a,Pereira09b,Guinea2010,Levy2010,Vozmediano10,Low2011}. 
We show that under an applied strain the bond vectors of 2d materials generally transform as 
$\vec{r}_{ij}\to (\ten{I}_3+\ten{\strain})\cdot\vec{r}_{ij} + \vec{\Delta}_\parallel + \Delta_\perp\hat{z}$. 
The vectors $\vec{\Delta}_\parallel$ and $\Delta_\perp\hat{z}$ are determined by minimizing 
the deformation energy. For graphene, the strain-displacement effect on the electronic 
properties is to renormalize the vector potential by a factor $(1-\kappa)\approx3/5$ while 
keeping its functional form. We provide a simple estimate of this correction based on 
the macroscopically observable Poisson ratio $\nu$, as $(1-\kappa)\approx(1-\nu)/(1+\nu)$.
The $(1-\kappa)$ renormalization alters the dependence of the band gap in AGNRs on strain: 
it increases the strain required to reach a certain band gap. Having a well controlled band gap 
is of key importance in a variety of applications\cite{Pereira09a,Pereira09b,caro+13,yacu+14,micr15}. 

For 2d materials with a more complex crystal structure than graphene, such as the TMDCs 
and phosphorene, the strain-displacement relations we put forward are key to understand 
their fundamental material properties. For phosphorene, we show (see Supplementary Material) that 
its anisotropic mechanical properties\cite{wepe14,jira+15} can only be obtained by considering 
proper strain-displacement relations. 
Furthermore, phosphorene is considered to be very interesting 
from a strain-engineering 
perspective as it displays a strain dependent band-gap\cite{ximo+14,pequ+14,feya14}. Our strain-displacement
relations lead to an anisotropic renormalization of this gap.
This Letter provides a transparent approach to estimate the electromechanical coupling in 2d materials, 
given that the mechanical and electrical properties are reliably parameterized.

\begin{acknowledgments}
    The authors thank Andreas Isacsson and Nuno Peres for helpful comments.
\end{acknowledgments}


\providecommand{\latin}[1]{#1}
\providecommand*\mcitethebibliography{\thebibliography}
\csname @ifundefined\endcsname{endmcitethebibliography}
  {\let\endmcitethebibliography\endthebibliography}{}

\clearpage
\widetext
\begin{center}
\textbf{\large Supplemental Materials: Strain-displacement relations and strain engineering in 2d materials}
\end{center}
\setcounter{equation}{0}
\setcounter{figure}{0}
\setcounter{table}{0}
\setcounter{page}{1}
\makeatletter
\renewcommand{\theequation}{S\arabic{equation}}
\renewcommand{\thefigure}{S\arabic{figure}}
\renewcommand{\thetable}{S\arabic{table}}
\renewcommand{\bibnumfmt}[1]{[S#1]}
\renewcommand{\citenumfont}[1]{S#1}

\section*{Strain-displacement relations for black phosphorous}

\begin{figure*}[b!]
  \centering
  \includegraphics[width=0.5\textwidth]
             {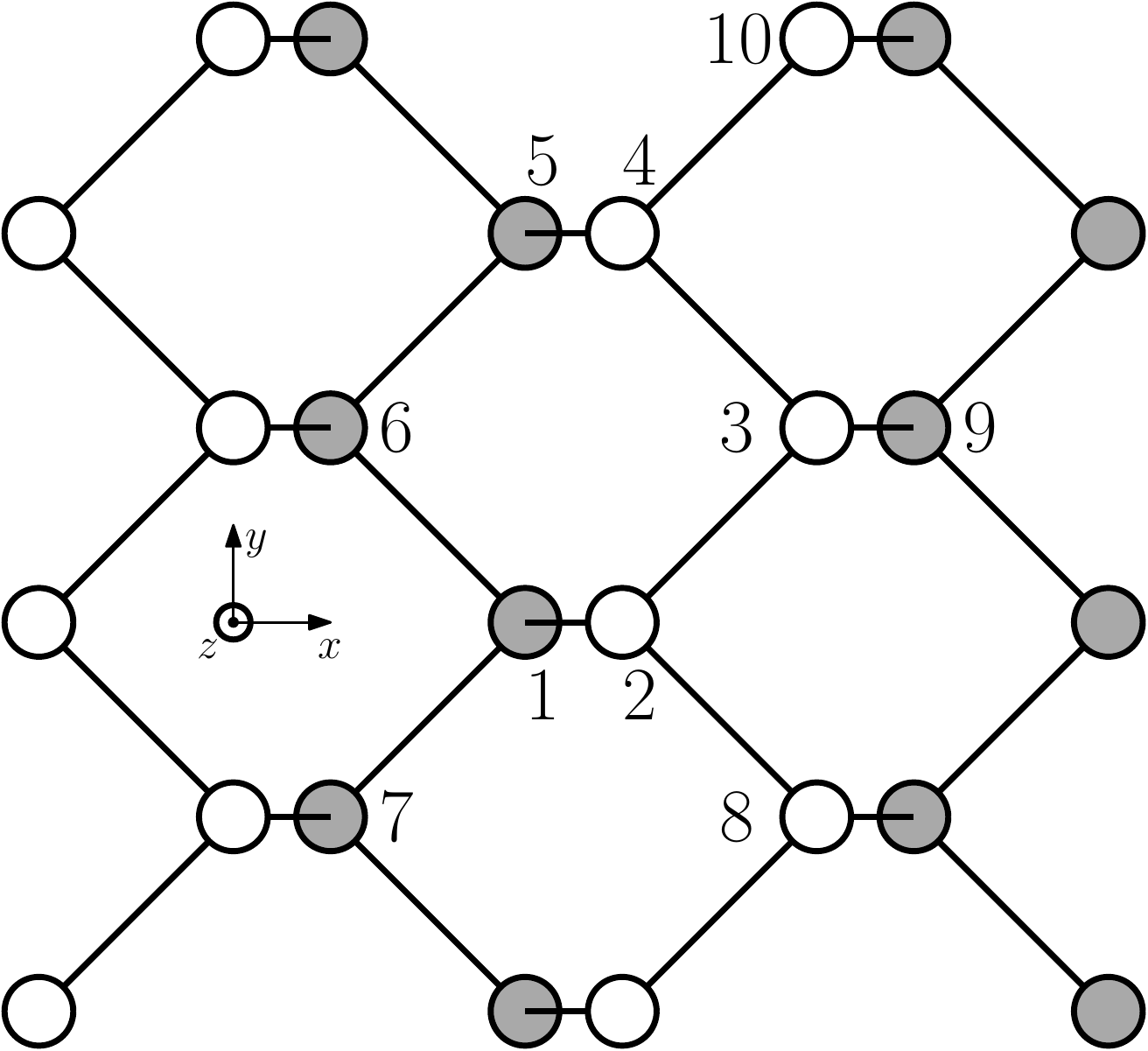}           
             \caption{Lattice structures of phosphorene. It is characterized by two lattice vectors and four basis atoms ($1,2,3,4$), defining ten nearest neighbor vectors.}
\label{fig:Lat}
\end{figure*}
Black phosphorous is a two-dimensional material with an orthorhombic lattice-structure with lattice vectors $\vec{a}_1$ and $\vec{a}_2$, and four basis atoms arranged in a puckered structure (see Fig.\ \ref{fig:Lat}). The bond vectors have approximately equal length $d\approx 2.24$ \AA, and intra- and inter-pucker angles $\Theta_1 =\Theta_{328}\approx 96.3^{\rm o}$ and $\Theta_2=\Theta_{123} \approx 102^{\rm o}$  \cite{ji15a}. Hence, the bond vectors read
\begin{subequations}\label{eq:r0}
\begin{align}
    \vec{r}_{12} ={}& [-d \cos(\Theta_2)/\cos(\Theta_1/2), 0, d \sqrt{1- \left(\cos(\Theta_2)/\cos(\Theta_1/2)\right)^2} ]\;,\\
    \vec{r}_{23} ={}& [ d \cos(\Theta_1/2), d \sin(\Theta_1/2), 0]\;,\\
    \vec{r}_{43} ={}& [ d \cos(\Theta_1/2), -d \sin(\Theta_1/2), 0]\;.
\end{align}
\end{subequations}
with $\vec{r}_{43}=\vec{r}_{28}=\vec{r}_{61}$, 
$\vec{r}_{23}=\vec{r}_{41}=\vec{r}_{71}$, 
$\vec{r}_{12}=\vec{r}_{54}$
and $\vec{r}_{39}=[r_{12,x}, r_{12,y}, -r_{12,z}]$.
A lattice deformation, cast in terms of the strain tensor $\ten{\strain}$, modifies the bond vectors as
\begin{subequations}\label{eq:rp}
\begin{align}
    \vec{r}'_{12} ={}& \vec{r}_{12} + \ten{\strain}\cdot \vec{r}_{12} + \vec{\Delta}_\parallel + \vec{\Delta}_\perp \hat{z} \;,\\
    \vec{r}'_{23} ={}& \vec{r}_{23} + \ten{\strain}\cdot \vec{r}_{23} - \vec{\Delta}_\parallel \;,\\
    \vec{r}'_{43} ={}& \vec{r}_{43} + \ten{\strain}\cdot \vec{r}_{43} - \vec{\Delta}_\parallel \;.
\end{align}
\end{subequations}
We assume that $\vec{r}'_{43}=\vec{r}'_{28}=\vec{r}'_{61}$, 
$\vec{r}'_{23}=\vec{r}'_{41}=\vec{r}'_{71}$, 
$\vec{r}'_{12}=\vec{r}'_{54}$
and $\vec{r}'_{39}=[r'_{12,x}, r'_{12,y}, -r'_{12,z}]$.

According to the VFM put forward in Ref.\ \onlinecite{kaka+82} 
(see also Ref.\ \onlinecite{ji15a}) the elastic energy
of the unit cell is given by
\begin{align}
    E_{\rm cell} ={}& K_r d^2\, \left( \delta r_{23} \right)^2 + K_r d^2\, \left( \delta r_{28} \right)^2 + K'_{r}d^2\, \left( \delta r_{21} \right)^2 \notag\\
                    &+ 2 K_{\Theta}d^2\, \left( \delta \Theta_{328} \right)^2
        + 2 K'_{\Theta}d^2\, \left( \delta \Theta_{321} \right)^2
        + 2 K'_{\Theta}d^2\, \left( \delta \Theta_{821} \right)^2 \notag\\
        &+ 2 K_{rr'}d^2\, \left( \delta r_{23} \delta r_{28} \right)
        + 2 K'_{rr'}d^2\, \left( \delta r_{23} \delta r_{21} \right)
        + 2 K'_{rr'}d^2\, \left( \delta r_{28} \delta r_{21} \right) \notag\\
        &+ 2 K_{r\Theta}d^2\, \left( \delta r_{23} \delta \Theta_{234} \right)
        + 2 K_{r\Theta}d^2\, \left( \delta r_{43} \delta \Theta_{234} \right)
        + 2 K'_{r\Theta}d^2\, \left( \delta r_{23} \delta \Theta_{123} \right)
        + 2 K'_{r\Theta}d^2\, \left( \delta r_{28} \delta \Theta_{128} \right) \notag\\
        &+ 2 K''_{r\Theta}d^2\, \left( \delta r_{12} \delta \Theta_{123} \right)
        + 2 K''_{r\Theta}d^2\, \left( \delta r_{12} \delta \Theta_{128} \right)\;,\label{eq:en_cell}
\end{align}
where $\delta r_{ij}=|\vec{r}'_{ij} - \vec{r}_{ij}|/d \approx (\vec{r}'_{ij} - \vec{r}_{ij})\cdot\vec{r}_{ij}/d^2$ 
is the relative bond-length change and 
$\delta \Theta_{ijk} \approx - (\cos(\Theta_{ijk}' )- \cos(\Theta_{ijk}))/\sin(\Theta_{ijk})$ 
with $\cos(\Theta_{ijk}')\approx (\vec{r}'_{ij}\cdot \vec{r}_{jk}')(1-\delta r_{ij}/d-\delta r_{jk}/d)/d^2$ 
is the bond angle change due to the deformation up to first order in strain and $\vec{\Delta}$. 
There are nine force-field parameters, namely, $K_r$, $K'_r$,  $K_{\Theta}$, $K'_{\Theta}$, 
$K_{rr'}$, $K_{r\Theta}$, $K'_{r\Theta}$ and $K''_{r\Theta}$. 
Those determine the energy cost for bond stretching, angle bending, bond-bond and 
bond-angle correlations.
The expressions for the bond vectors, Eqs.\ \eqref{eq:r0} and \eqref{eq:rp}, render 
the energy $E_{\rm cell}$, Eq. \eqref{eq:en_cell}, to be a quadratic function in the components 
of the strain 
tensor and the components of $\vec{\Delta}$. 
By minimizing $E_{\rm cell}$ with respect to $\Delta_{\parallel,x}$, $\Delta_{\parallel,y}$ 
and $\Delta_{\perp}$ and dividing by the area of the unit cell, $A_{\rm cell}$, 
we obtain the deformation energy-density $\mathcal{E}_{\rm cell}=E_{\rm cell}/A_{\rm cell}$.
For the VFM parameters reported in Ref.\ \onlinecite{kaka+82} we find 
\begin{equation}\label{eq:delta}
    \Delta_{\parallel, x} = 0.61 d\; \strain_{xx} + 0.37 d\; \strain_{yy} \;,\quad
    \Delta_{\parallel, y} = 1.02 d\; \strain_{xy} \;,\quad
    \Delta_{\perp} = -0.36 d\; \strain_{xx} - 0.20 d\; \strain_{yy}\;.
\end{equation}
These expressions, together with Eqs.\ \eqref{eq:rp}, constitute the strain-displacement relations for
phosphorene.

\subsection{Elasticity constants}

Having the deformation energy-density $\mathcal{E}_{\rm cell}$ one can calculate
the elastic constants $C_{ij}$, which are given by\cite{wepe14}
\begin{equation}\label{eq:elastic}
    C_{ij} = \frac{\partial^2 \mathcal{E}_{\rm cell}}{\partial \strain_i \partial \strain_j}\;,
\end{equation}
where $i,j=1,2,6$ and $\strain_1=\strain_{xx}$, $\strain_2=\strain_{xx}$ and $\strain_6=\strain_{xy}$.
From those one derives Young's and shear moduli, $Y_{x/y}$ and $G_{xy}$, and the Poisson ratios
$\nu_{xy/yx}$,
\begin{equation}\label{eq:young}
    Y_x = \frac{C_{11}C_{22}-C_{12}^2}{C_{22}}\;,\quad
    Y_y = \frac{C_{11}C_{22}-C_{12}^2}{C_{11}}\;,\quad
    G_{xy} = C_{66}\;,\quad
    \nu_{xy} = \frac{C_{12}}{C_{22}}\;,\quad
    \nu_{yx} = \frac{C_{12}}{C_{11}}\;.
\end{equation}
Moreover, the sound velocities are given by\cite{lali86}
\begin{equation}\label{eq:sound}
    v_{xx} = \sqrt{\frac{C_{11}}{\rho_P}}\;,\quad
    v_{xx} = \sqrt{\frac{C_{22}}{\rho_P}}\;,\quad
    v_{xy} = \sqrt{\frac{C_{66}}{\rho_P}}\;,
\end{equation}
where $\rho_P$ is the 2d mass density of phosphorene.

\begin{table}[tb]
\begin{tabular}{lccc}
                        & $Y_x$ & $Y_y$ & $G_{xy}$ \\
    \hline
without correction      & $58$  & $95$  & $51$\\
with correction         & $17$  & $94$  & $20$\\
\hline
\hline
                        & $v_{xx}$ & $v_{yy}$ & $v_{xy}$ \\
    \hline
without correction      & $6993$   & $8930$   & $5957$ \\
with correction         & $3508$   & $8147$   & $3707$ \\
Kaneta et al. \cite{kaka+82}          & $3720$   & $8190$   & $3760$ 
\end{tabular}
\caption{Young's and shear moduli and sound velocities from Eqs.\ \eqref{eq:young} and \eqref{eq:sound}. The former are given
in units of $N/m$ and the latter in $m/s$.}
\label{tab:el_res}
\end{table}
The results we calculate for all those elastic properties are summarized in Tab.\ \ref{tab:el_res}.
We also provide results for the case where the correction $\vec{\Delta}$ is not taken into
account. Comparing the sound velocities to the values given in Ref.\ \onlinecite{kaka+82}
one readily concludes that is necessary to account for  $\vec{\Delta}$ to get comparable results.
One may also compare the Young's moduli to recent DFT results from various groups. There the values of the reported Young's moduli are for instance: 
$26$ N/m and $88$ N/m \cite{elkh+15}, $24$ N/m and $92$ N/m \cite{wepe14}, $29$ N/m and $102$ N/m \cite{qiko+14}.
Again the values in Tab.\ \ref{tab:el_res} for the corrected case match best.

Qualitatively, the VFM reported in Ref.\ \onlinecite{kaka+82} gives very good results. However, it should also be mentioned that
it has some deficits. For example, the Poisson ratios are obtained as $\nu_{yx}=0.27$ and $\nu_{xy}=0.05$, whereas other
studies report $0.81/0.24$\cite{elkh+15} or $0.62/0.17$\cite{wepe14}. However, it should be kept in mind that the parameters
of the VFM were found by fitting to optical $\Gamma$-phonons measured in an experiment. To get a better quantitative description
more data is required and the Poisson ratios need to be included in the fitting procedure.

\subsection{Electronic band-gap}

As argued in the main text, the strain-displacement relations are very important for correctly estimating 
the influence of strain on the electronic structure. To demonstrate this for phosphorene, we follow
Ref.\ \onlinecite{jipa15} where a two-orbital tight-binding model was used to calculate the electronic
band-gap of phosphorene. The band-gap energy is then given by
\begin{equation}
    E_{\rm gap} = 2 (t_1 + t_2 + t_3)\;,
\end{equation}
where $t_1$, $t_2$ and $t_3$ are the hopping parameters between atoms $2$ and $3$, $2$ and $1$ and
$2$ and $8$, respectively. For the undeformed lattice the values are $t_1^0=t_3^0=-0.797$ eV and $t_2^0=2.393$ eV.
Assuming a distance dependence of the hopping parameters as $t(r)\propto 1/r^2$, in accordance with the nature
of the orbitals responsible for the relevant electronic bands, allows it to calculate the strain-induced
modification of the band-gap $\Delta E_{\rm gap}$. To this end we use the strain-displacement relations,
Eq.\ \eqref{eq:rp} and \eqref{eq:delta}, for the respective bond-lengths. We obtain
\begin{align}
    \text{without correction:}\qquad & \Delta E_{\rm gap} \approx 1.899 \strain_{xx} + 3.538 \strain_{yy}\;, \\
    \text{with correction:}\qquad & \Delta E_{\rm gap}    \approx 0.755 \strain_{xx} + 2.612 \strain_{yy}\;.
\end{align}
For comparison, in Ref.\ \onlinecite{jipa15} the band-gap modification $\Delta E_{\rm gap} \approx 1.863 \strain_{xx} + 3.507 \strain_{yy}$
was obtained by transforming all bond-vectors using the strain tensor only. Similar to the case of graphene, one obtains 
a non-negligible renormalization due to the non-Bravais nature of the lattice.


\providecommand{\latin}[1]{#1}
\providecommand*\mcitethebibliography{\thebibliography}
\csname @ifundefined\endcsname{endmcitethebibliography}
  {\let\endmcitethebibliography\endthebibliography}{}

\end{document}